\documentstyle[12pt]{article}

\newcommand{\beq}{\begin{equation}}
\newcommand{\eeq}{\end{equation}}
\newcommand{\ba}{\begin{array}}
\newcommand{\ea}{\end{array}} 
\newcommand{\beqa}{\begin{eqnarray}}
\newcommand{\eeqa}{\end{eqnarray}}
\newcommand{\dis}{\displaystyle}

\newcommand{\cH}{{\cal H}}

\newcommand{\cO}{{\cal O}}

\newcommand{\da}{^\dagger}
\newcommand{\no}{\nonumber}
\newcommand{\Frac}[2]{\frac{\displaystyle #1}{\displaystyle #2}}

\newcommand{\Ko}{{K}^0}
\newcommand{\Kob}{\bar{K}^0}

\newcommand{\eps}{\epsilon}
\newcommand{\epsp}{\epsilon'}    
\newcommand{\epop}{\frac{\epsilon'}{\epsilon}}

\newcommand{\tq}{{\tilde q}}

\newcommand{\td}{{\tilde d}}
\newcommand{\ts}{{\tilde s}}
\newcommand{\tg}{{\tilde g}}

\renewcommand{\Im}{\mbox{Im}}
\renewcommand{\Re}{\mbox{Re}}
\newcommand{\SUSY}{{\mbox{\tiny SUSY}}}

\newcommand{\PL}[3]{{\it Phys.\ Lett.\ }         {\bf #1}, {#3} {(19#2)}}
\newcommand{\PRe}[3]{{\it Phys.\ Rep.\ }         {\bf #1}, {#3} {(19#2)}}
\newcommand{\PRL}[3]{{\it Phys.\ Rev.\ Lett.\ }  {\bf #1}, {#3} {(19#2)}}
\newcommand{\PR}[3]{{\it Phys.\ Rev.\ }          {\bf #1}, {#3} {(19#2)}}
\newcommand{\NP}[3]{{\it Nucl.\ Phys.\ }         {\bf #1}, {#3} {(19#2)}}
\newcommand{\ZP}[3]{{\it Z.\ Phys.\ }            {\bf #1}, {#3} {(19#2)}}
\newcommand{\JHEP}[3]{{\it JHEP\ }               {\bf #1}, {#3} {(19#2)}}
\newcommand{\IJMP}[3]{{\it Int. J. Mod. Phys.\ } {\bf #1}, {#3} {(19#2)}}

\newcommand{\EPJ}[3]{{\it Eur.\  Phys.\  J.\  }  {\bf #1}, {#3} {(19#2)}}

\begin{document}
\begin{titlepage}
\begin{flushright} 
LNF--99/023(P) \\
FTUV/99--53 \\
IFIC/99--55 \\
ZU--TH 21/99
\end{flushright}
\begin{center}

\vspace*{1cm}
 
{\large \bf Supersymmetric contributions to \\
\vspace*{0.3cm}
direct $CP$ violation in $K \rightarrow \pi\pi\gamma$
decays$^*$}
 
\vspace*{1.4cm}
{\bf{G. Colangelo$^1$, G. Isidori$^2$ and J. Portol\'es$^3$}}\\[8pt]

${}^{1)}$ Institut f\"ur Theoretische Physik der Universit\"at Z\"urich,\\
Winterthurerstr. 190, CH--8057 Z\"urich--Irchel, Switzerland\\[5pt]

${}^{2)}$ INFN, Laboratori Nazionali di Frascati\\
P.O. Box 13, I--00044 Frascati, Italy\\[5pt]

${}^{3)}$ Departament de F\'{\i}sica Te\`orica (I.F.I.C.), \\
C.S.I.C.--Universitat de Val\`encia, Apdo. 2085,
E--46071 Val\`encia, Spain\\[10pt]

\vspace*{1.2cm}

{\bf Abstract} \\
\end{center}
We analyze the supersymmetric contributions to direct--$CP$--violating
observables in $K \to \pi\pi\gamma$ decays induced by gluino--mediated
magnetic--penguin operators.  We find that $\epsp_{+-\gamma}$ and the
differential width asymmetry of $K^\pm \to \pi^\pm \pi^0 \gamma$ decays
could be substantially enhanced with respect to their Standard Model
values, especially in the scenario where $\epsp/\eps$ is dominated by
supersymmetric contributions.  These observables could therefore provide a
useful tool to search for New Physics effects in $|\Delta S|=1$
transitions, complementary to $\epsp/\eps$ and rare decays.

\vfill
\noindent
{\footnotesize $^*$ Work supported in part by TMR, EC--Contract No. ERBFM
RX--CT980169.}

\end{titlepage}

\section{Introduction}
The phenomenon of $CP$ violation is one of the least tested aspects of the
Standard Model (SM) and represents one of the sectors where a large
sensitivity to possible New Physics (NP) effects can be expected.  An
important step forward in understanding the nature of this phenomenon has
recently been achieved by the KTeV and NA48 collaborations, obtaining the
following measurements of direct $CP$ violation in $\Ko(\Kob) \to 2 \pi$
decays: 
\beq 
\Re\left(\epop \right) = \left\{ \ba{ll} (28.0 \pm 4.1 )
\times 10^{-4} \; \; \; \; \; \; \; & \cite{KTeV}~, \\ (18.5 \pm 7.3)
\times 10^{-4} & \cite{sozzi}~. \ea \right.
\label{epspKTEV}
\eeq These results, together with the earlier finding by NA31 \cite{NA31},
clearly establish the existence of direct $CP$ violation, as generally
predicted by the SM. However, an intriguing aspect of this new measurement
is that the values in (\ref{epspKTEV}) tend to be larger than most SM
estimates \cite{Bosch,epsp2}.  Unfortunately the 
theoretical predictions of $\epsp/\eps$ are affected by large
uncertainties, mainly of non--perturbative origin, and it is possible that
the experimental values above are still compatible with the SM expectations
(see, in particular, Ref.~\cite{epsp2}). Nonetheless, it is clear that after
these new experimental results the chances of sizable NP contributions in
$\epsp/\eps$ have increased substantially.

Among other possible NP scenarios, low energy supersymmetry \cite{Susy}
represents one of the most interesting and consistent extensions of the
Standard Model.  In generic supersymmetric models, the large number of
new particles carrying flavor quantum numbers would naturally lead to large
effects in $CP$--violating and flavor--changing neutral--current (FCNC)
amplitudes \cite{SusyFCNC,HKR}.  Actually, in this context the problem is
not how to generate large $CP$--violating effects, but rather how to avoid
dangerous corrections to small quantities like $\epsilon_K$ or $\Delta
m_K$, which seem to be consistent with their SM expectations. However, as
discussed recently in \cite{Sanda,Masiero,noi}, in specific supersymmetric
scenarios it is possible to generate non--standard $\cO(10^{-3})$
contributions to $\epsp/\eps$ without getting troubles with the
experimental constraints of other $CP$ and FCNC processes.
\par
From a phenomenological point of view, the supersymmetric sources of a
sizable enhancement of $\epsp/\eps$ which can avoid fine--tuning problems
in $|\Delta S|=2$ amplitudes, are basically two \cite{noi}: a large
$\bar{s}dG$ vertex induced by the chromomagnetic operator \cite{Masiero}
and an enhanced $\bar{s} d Z$ vertex \cite{CI}. Since the problem of
non--perturbative uncertainties in the estimate of $\epsp/\eps$ is
typically worse in the case of supersymmetric contributions, it is very
useful to identify other observables which could clearly signal the
manifestation of either of these two mechanisms.  As discussed in
\cite{noi,BS}, in the case of the enhanced $\bar{s} d Z$ vertex there is a
strong correlation between $\epsp/\eps$ and the theoretically--clean
$K\to\pi\nu\bar{\nu}$ widths. The scenario where $\epsp/\eps$ receives
sizable supersymmetric corrections via the $\bar{s} d Z$ vertex could
therefore be clearly excluded or confirmed by future precise experiments on
rare decays.
\par
More difficult to identify is the case where $\epsp/\eps$ receives
sizable contributions by the chromomagnetic operator. Indeed this
non--standard effect would be present mainly in non--leptonic
processes. However, since there is a strict correlation between the
chromomagnetic operator ($\sim \bar{s} \sigma^{\mu\nu} t^a d G^a_{\mu\nu}
$) and the magnetic penguin contributing to the $s\to d \gamma$ transition
($\sim \bar{s} \sigma^{\mu\nu} d F_{\mu\nu}$), interesting consequences of
this scenario could in principle be observed in processes with real photons
or $e^+e^-$ pairs in the final state.  As shown in \cite{noi}, an example
of such processes is provided by the $K_L \to \pi^0 e^+e^-$ decay.  In this
letter we analyze the consequences of this scenario in $K\to\pi\pi \gamma$
decays, focusing on the possible enhancements of direct--$CP$--violating
observables.  As we will show, these can provide complementary information
to rare decays.

The paper is organized as follows: in Section 2 we recall the structure of
supersymmetric contributions to magnetic operators and their impact on
$\epsp/\eps$. In Section 3 we estimate the matrix element of the tensor
current, necessary to evaluate $CP$--violating effects in $K\to\pi\pi
\gamma$ decays.  The general decomposition of $K\to\pi\pi \gamma$
amplitudes and the estimate of the supersymmetric contributions to
$\epsp_{+-\gamma}$ is given in Section 4, while in Section 5 we discuss the
charge asymmetry in $K^\pm\to\pi^\pm \pi^0 \gamma$ decays. Finally in
Section 6 we summarize our results.

\section{Gluino contributions to magnetic operators
 and $\epsp/\eps$} 
A useful framework to evaluate supersymmetric contributions to
$CP$--vio\-la\-ting and FCNC processes is provided by the mass--insertion
approximation \cite{HKR}. This consists in choosing a simple flavor--basis
for the gauge interactions and, in that basis, to perform a perturbative
expansion of the squark mass matrices around their
diagonal. Gluino--mediated amplitudes usually provide the dominant effect,
therefore the basis typically adopted is the one where the
gluino--quark--squark vertices are flavor--diagonal.

A detailed discussion of the leading terms generated by gluino exchange in
the framework of the mass--insertion approximation can be found in
\cite{GGMS,CFGMS}.  Given the strong constraints from $|\Delta S|=2$
processes, it is found that only the dimension--5 magnetic operators
induced by $\td_{L(R)}-\ts_{R(L)}$ mixing could lead to sizable
$CP$--violating effects in $|\Delta S|=1$ amplitudes avoiding fine--tuning
problems.  These operators can be written as \cite{GGMS} 
\beqa
&&\cH^{(5)}_{eff}~=~\frac{(\delta_{RL}^D)_{21}}{m_\tg} \left[ {\widetilde
C}_7(x_{gq}) {\bar s}_R \sigma^{\mu\nu}d_L {\hat F}_{\mu\nu} + {\widetilde
C}_8(x_{gq}){\bar s}_R \sigma^{\mu\nu}{\hat G}_{\mu\nu} d_L \right] \no\\
&&\qquad + \frac{(\delta_{LR}^D)_{21}}{m_\tg} \left[ {\widetilde
C}_7(x_{gq}) {\bar s}_L \sigma^{\mu\nu}d_R {\hat F}_{\mu\nu} + {\widetilde
C}_8(x_{gq}) {\bar s}_L \sigma^{\mu\nu} {\hat G}_{\mu\nu} d_R \right]+{\rm
h.c.}~,\qquad
\label{Heff}
\eeqa
where ${\hat G}_{\mu\nu}=gt^a G^a_{\mu\nu}$, ${\hat F}_{\mu\nu}=e F_{\mu\nu}$, 
\beq
 (\delta_{AB}^D)_{ij} = (\delta_{BA}^D)_{ji}^* = (M^2_D)_{\tq^i_A \tq^j_B}/m^2_{\td}~, 
\eeq
$m_{\td}$ is the average down--squark mass, 
$m_{\tg}$ is the gluino mass and $x_{gq}=m_{\tg}^2/m^2_{\td}$.
Neglecting QCD corrections, the Wilson coefficients 
${\widetilde C}_{7,8}(x_{gq})$ are given by \cite{noi,GGMS}
\beqa
&{\widetilde C}_7 (x) = -\dis\frac{\alpha_s}{24\pi} F_0(x)~,  \qquad\qquad
&{\widetilde C}_7 (1) = -\dis\frac{1}{108} \frac{\alpha_s}{\pi}~, \\
&{\widetilde C}_8 (x) = \dis\frac{\alpha_s}{8\pi} G_0(x)~, \qquad\quad \qquad 
&{\widetilde C}_8 (1) = -\dis\frac{5}{144} \frac{\alpha_s}{\pi}~,
\eeqa
with 
\beqa
G_0(x) &=& \frac{x(22-20x-2x^2+16x\ln(x)
  -x^2\ln(x)+9\ln(x))}{3(1-x)^4}~, \\
F_0(x) &=& \frac{4x(1+4x-5x^2+4x\ln(x)
  +2x^2\ln(x))}{3 (1-x)^4}~.
\eeqa

Due to the smallness of the electric charge, the contribution generated by
$\cH^{(5)}_{eff}$ to $\Re(\epsp/\eps)$ is dominated by the terms
proportional to ${\widetilde C}_8$.  This can be written as \cite{noi} 
\beq
\Re \left(\frac{\epsp}{\eps}\right)^\SUSY_G = P_G \Im \Lambda^-_g ~,
\label{epspG} \eeq where \beq \Lambda^-_g = \left[ (\delta_{LR}^D)_{21} -
(\delta_{LR}^D)_{12}^* \right] G_0(x_{gq}) 
\eeq 
and\footnote{~Following
\cite{DAI}, here we adopt a normalization of $K\to (2 \pi)_I$ amplitudes
such that $\Re(A_0)^{\rm exp}=2.72\times 10^{-7}$~GeV and we employ the
notation $F_\pi = 92.4$~MeV.  Note that both these conventions differ from
those adopted in \protect\cite{noi}. Moreover $\omega^{-1} =
(\Re(A_0)/\Re(A_2))_{exp} = 22.2 \pm 0.1$ is the $\Delta I = 1/2$ rule
enhancement factor.}
\beqa  
P_G~ & = & ~\frac{11}{64}
\frac{\omega}{|\eps|\Re(A_0)} \frac{ m_\pi^2 m_K^2}{F_\pi (m_s+m_d)}
\frac{\alpha_s(m_\tg)}{\pi} \frac{1}{m_\tg} \eta B_G \no \\ & \simeq &
2.4\times 10^{2} B_G \left(\frac{137 {\rm~MeV}}{m_s + m_d}\right)
\left(\frac{ 500~\mbox{GeV}}{m_\tg}\right) 
\left(\frac{\alpha_s(m_\tg)}{\alpha_s(500~{\rm
      GeV})}\right)^{\frac{23}{21}}~. \qquad \label{PG}  
\eeqa 
The expression (\ref{epspG}) has been obtained neglecting the mixing 
induced by QCD corrections between ${\widetilde C}_8$ and the Wilson 
coefficients of the SM $|\Delta S|=1$ effective Hamiltonian.  This is a good approximation if
${\widetilde C}_8$ is sufficiently large: in this case the
renormalization--group evolution of ${\widetilde C}_8$ is almost diagonal
and is taken into account by the factor \cite{noi} 
\beq 
\eta =
\left(\frac{\alpha_s(m_\tg)}{\alpha_s(m_t)}\right)^\frac{2}{21}
\left(\frac{\alpha_s(m_t)}{\alpha_s(m_b)}\right)^\frac{2}{23}
\left(\frac{\alpha_s(m_b)}{\alpha_s(m_c)}\right)^\frac{2}{25} \simeq 0.89
\left(\frac{\alpha_s(m_\tg)}{\alpha_s(500~\rm{GeV})} \right)^\frac{2}{21}~.
\eeq 
In (\ref{PG}) we have not explicitly shown the scale dependence of
quark masses and $B_G$, which are evaluated at $\mu = m_c$.  The parameter
$B_G$, expected to be $\cO(1)$ for a renormalization scale $\mu \sim
1$~GeV, is defined by 
\beq 
\langle (\pi\pi)_{I=0} | {\bar s}_R
\sigma^{\mu\nu}{\hat G}_{\mu\nu} d_L |K^0 \rangle (\mu) =
\frac{11}{4\sqrt{2}} \frac{m_\pi^2}{F_\pi}\frac{m_K^2}{m_s(\mu)+m_d(\mu)}
B_G(\mu)~.  
\eeq

\section{Matrix elements of the tensor current}
Contrary to the case of $\epsp/\eps$, the ${\widetilde C}_7$ terms of
$\cH^{(5)}_{eff}$ could play an important role in $CP$--violating
observables of $K\to\pi\pi\gamma$ decays.  In order to evaluate their
impact, we need to estimate the matrix elements of the
$\bar{s}_{R(L)}\sigma^{\mu\nu} d_{L(R)}$ current between kaon and pion
states.  Given the Lorentz structure and the transformation properties
under $CP$ and $SU(3)_L\times SU(3)_R$, the lowest--order chiral
realization of the tensor current can be written as
\beqa
\bar{s}_R \sigma_{\mu\nu} d_L &\longrightarrow &
-i \frac{a_T F_\pi^2}{2} \left[ \partial_\mu U\da \partial_\nu U U\da - 
 \partial_\nu U\da \partial_\mu U U\da\right]_{23}~, \label{tc1}\\
\bar{s}_L \sigma_{\mu\nu} d_R &\longrightarrow &
-i \frac{a_T F_\pi^2}{2} \left[ \partial_\mu U \partial_\nu U\da U - 
 \partial_\nu U \partial_\mu U\da U \right]_{23}~, \label{tc2}
\eeqa 
where we have neglected terms proportional to the Levi-Civita tensor,
$\eps_{\mu\nu\rho\sigma}$, not interesting to the present analysis.
Here $U$ is the usual chiral field (we follow the notation 
of \cite{DAI}) and $a_T$ is an unknown coupling. 

To obtain a first estimate of $a_T$ we proceed by differentiating 
and using the e.o.m. on both sides of (\ref{tc1}-\ref{tc2}).
In this way on the l.h.s. we obtain some terms whose
chiral realization is well known, namely the 
$\bar s_{L(R)} \gamma^\mu d_{L(R)}$ currents. Identifying the 
corresponding terms on the r.h.s. we then obtain
\beq
a_T= \frac{m_s+m_d}{m_K^2}~.
\label{at1}
\eeq 
Unfortunately, it is not possible to repeat this identification for
all the quark bilinears which appear on the l.h.s.  This shows that
Eq. (\ref{at1}) is not to be trusted literally. The same conclusion can
also be reached by noting that the scale dependence of the tensor current
is not the same as that of the scalar bilinear. Eq. (\ref{at1}) would
therefore give the wrong scale dependence of the matrix elements of the
tensor current, and, strictly speaking, cannot be correct.
On the other hand, we find Eq. (\ref{at1}) instructive, in
the sense that it shows that the coefficient $a_T$ (which has dimensions of 
the inverse of a mass) must be proportional to the inverse of the scale of
chiral symmetry breaking, with a numerical coefficient of $\cO(1)$.

An additional indication on the value of $a_T$ can be obtained by
evaluating the $\langle K | \bar{s} \sigma^{\mu\nu} d | \pi \rangle$ matrix
element in the limit where the strange quark mass is very heavy ($m_s \gg
\Lambda_{QCD}$).  The value of $a_T$ thus determined can be written as
\cite{Casalbuoni} 
\beq 
|a_T| \simeq \frac{1}{2 m_K}\left[ f_+(q^2)
+\cO(f_-) \right]~,
\label{at2}
\eeq 
where $f_\pm (q^2)$ are the form factors of the vector current.  
Obviously, this result can be
trusted even less than Eq. (\ref{at1}). On the other hand it shows that if
we vary the strange quark mass, and approach its physical value from above, 
we get a value of $a_T$ which is numerically close to that obtained with
chiral arguments. We believe that this serves as an independent check of 
the order of magnitude, and gives us confidence that the real value of 
$a_T$ cannot be too different from the estimates presented here. A further 
independent estimate of $|a_T|$ very close to the one in (\ref{at2}) can be 
obtained also in the framework of vector meson dominance, as in 
\cite{RPS}. Given these results, for simplicity we shall assume in the
following 
\beq 
a_T = \frac{B_T}{2 m_K}~,
\label{at3}
\eeq
where $B_T$ is a dimensionless parameter expected to be of $\cO(1)$. Note, 
however, that Eq. (\ref{at3}) does not show the correct chiral behaviour,
which should rather be read from (\ref{at1}). Both the correct dependence
on the quark masses, and on the QCD renormalization scale are assumed to be 
hidden inside $B_T$.

\section{$K \to \pi \pi \gamma$ amplitudes and $\epsilon'_{+-\gamma}$}
The most general form, dictated by gauge and Lorentz invariance, for the
transition amplitude $K(p_K) \to \pi_1(p_1)\pi_2(p_2)\gamma(\epsilon, q)$
is given by \beq A(K\to\pi\pi\gamma)= \epsilon_{\mu}^* \left[ E(z_i)
(qp_1p_2^\mu - qp_2p_1^\mu) + M(z_i)
\epsilon^{\mu\nu\rho\sigma}p_{1\nu}p_{2\rho}q_{\sigma}\right] /m_K^3,
\label{kppgamp1} \eeq where $E$ and $M$, known as electric and magnetic
amplitudes, are dimensionless functions of 
\beq 
z_i = {p_iq \over m_K^2} \quad(i=1,2)\qquad \mbox{and} \qquad z_3 = z_1+z_2
= {p_K q \over m_K^2} 
\eeq 
(only two of the $z_i$'s are independent).  Following \cite{DAI} we
can decompose the electric amplitude as $E = E_{IB} + E_{DE}$, where 
\beq
E_{IB}(z_i) = { e A(K\to \pi_1\pi_2) \over M_K z_3 }\left ( {Q_2 \over z_2}
- {Q_1 \over z_1} \right) 
\eeq 
is the well--known bremsstrahlung contribution ($eQ_i$ denotes the electric
charge of the pion $\pi_i$).  Furthermore, we can expand the
direct--emission amplitudes $E_{DE}$ and $M$ as

\beqa
E_{DE}(z_i) &=& E_1 + O\left[(z_1-z_2)\right]~, \\
M(z_i) &=& M_1 + O\left[(z_1-z_2)\right]~,
\label{eq:multi}
\eeqa 
where the higher order terms in $(z_1-z_2)$ can be safely neglected
due to the phase--space suppression.

The first $CP$ violating observable we shall consider is 
\beq
\eta_{+-\gamma} = { A (K_L \to \pi^+\pi^-\gamma)_{E_{IB} +E_1} \over
A (K_S \to \pi^+\pi^-\gamma)_{E_{IB} +E_1} }~.
\label{eta+-g} 
\eeq    
Due to the vanishing of direct emission amplitudes, at small photon
energies $\eta_{+-\gamma}$ tends to the usual $K\to 2\pi$ parameter
$\eta_{+-} = A(K_L \to \pi^+\pi^-)/A (K_S \to \pi^+\pi^-)$.  On the other
hand, the difference $(\eta_{+-\gamma}-\eta_{+-})$, that vanishes for
$E_\gamma \to 0$, is an independent index of direct $CP$
violation. Following \cite{DAI} we can write
\beq
\epsilon'_{+-\gamma}~=~\eta_{+-\gamma}-\eta_{+-}~=~
  i \frac { e^{i(\delta_n-\delta_0)} m_K z_+z_-}{e \sqrt{2} \Re A_0} 
  \left( \Im A_0 {\Re E_n \over \Re A_0 } - \Im E_n \right)~,
\label{eppmg}
\eeq
where on the r.h.s we have neglected small contributions suppressed by
$\omega = \Re A_2/\Re A_0 =0.045$ and the following decomposition has been
employed
\beq
E_1(K^0) = \frac{1}{\sqrt{2}}~e^{i\delta_n} E_n~, \qquad\qquad
(p_1,p_2)\equiv (p_+,p_-)~. 
\label{eq:e1en}
\eeq

Assuming that the dominant SUSY contribution to the $CP$--violating phase
of $E_n$ is generated by the magnetic photon operator we find
\beq
 \Im \left( E_n \right)^\SUSY  = 
  - \frac{em_K^2}{12 F_\pi}
 \frac{\alpha_s(m_\tg)}{\pi } \frac{ \eta^2
 B_T }{ m_\tg } \left[  \frac{ F_0(x_{gq}) }{ G_0(x_{gq}) }
       +  8 (1- \eta^{-1})  \right]
    \Im \Lambda^-_g~.
\label{eq:imen}
\eeq
Then using (\ref{epspG}) to express both $\Im \Lambda^-_g$ and $(\Im
A_0)^\SUSY_G$ in terms of $\Re(\epsp/\eps)^\SUSY_G$, we obtain
\beq
\left( \frac{\epsilon'_{+-\gamma}}{\eps}\right)^\SUSY = 
  \frac { e^{i(\delta_n-\delta_0+\pi/4)}  z_+z_-}{\omega} \left[
  R_{FG} - \frac{m_K \Re E_n}{e \Re A_0}  \right] 
  \Re \left(\frac{\epsp}{\eps}\right)^\SUSY_G~,
\label{eppmg2}
\eeq 
where 
\beqa 
R_{FG}~& = & \frac{16}{33\sqrt{2}}
\frac{m_K(m_s+m_d)}{m_\pi^2} \eta \frac{B_T}{B_G} \left[ \frac{F_0(x_{gq})
}{G_0(x_{gq})} +8 (1- \eta^{-1})\right]\qquad \label{RFG} \\ & \simeq & -
1.9 \frac{B_T}{B_G} \left( \frac{ m_s + m_d }{ 137 {\rm~MeV} } \right)
\qquad ({\rm for}\ m_\tg=500~\mbox{GeV}, x_{gq}=1)~. \no 
\eeqa
Unfortunately at the moment there are no precise experimental informations
about $\Re E_n$, however naive chiral counting suggests $m_K \Re E_n /( e
Re A_0) \ll 1$ \cite{DAI}. Neglecting this contribution in (\ref{eppmg2}),
assuming $|B_T/B_G|\leq 1$, $x_{gq}\leq 1.3$ \cite{RGEb} and $(m_s+m_d)\leq
158$~MeV, we finally obtain \beq \left| \frac{\epsilon'_{+-\gamma}}{\eps}
\right|^\SUSY \leq~50~z_+z_-~ \Re
\left(\frac{\epsp}{\eps}\right)^\SUSY_G\leq~0.15~ z_+z_-~,
\label{epgbound}
\eeq 
where the last inequality has been obtained imposing
$\Re(\epsp/\eps)^\SUSY_G \leq 3\times 10^{-3}$.  Note that the sensitivity
of this result to the value of $m_\tg$ and $m_\td$ 
is very small: they enter
only through the $F_0/G_0$ ratio and the factor $\eta$ in (\ref{RFG}).

Interestingly the upper bound (\ref{epgbound}) is substantially larger
(almost one order of magnitude) with respect to the corresponding one
obtained within the Standard Model \cite{DAI}.  A large value of
$\epsilon'_{+-\gamma}/\eps$ could therefore offer a clean signature of the
scenario where $\epsilon'/\eps$ is dominated by supersymmetric
magnetic--type contributions. Moreover, we notice that
$\epsilon'_{+-\gamma}/\eps$ is generated by the interference of two $\Delta
I = 1/2$ amplitudes (it is indeed enhanced by $\omega^{-1}$ with respect to
$\epsilon'/\eps$) and therefore, contrary to $\epsilon'/\eps$ or
$K_L\to\pi^0e^+e^-$, it is almost insensitive to possible new--physics
effects in the $\bar{s}dZ$ vertex.

Finally, we stress that the correlation between gluino--mediated
contributions to $\epsilon'/\eps$ and $\epsilon'_{+-\gamma}/\eps$ is
clearer than the corresponding one between $\epsilon'/\eps$ and
$B(K_L\to\pi^0e^+e^-)$ \cite{noi}.  Indeed, due to the different number of
pions in the final state, the supersymmetric coupling ruling the effect in
$K_L\to\pi^0e^+e^-$ is not exactly the same as in $\epsilon'/\eps$ and
$\epsilon'_{+-\gamma}/\eps$ \cite{noi}.

\section{Charge asymmetry in $K^\pm \to \pi^\pm \pi^0\gamma$}
A very clean observable of direct $CP$ violation is provided by the asymmetry
between $K^+ \rightarrow \pi^+ \pi^0 \gamma$ and $K^- \rightarrow \pi^-
\pi^0 \gamma$ decay widths \cite{DAI,RPS,CH67,DMS,DP}.  The decay rates of
$K^\pm \rightarrow \pi^\pm \pi^0 \gamma$ are conveniently expressed in
terms of $T_c^*$, the kinetic energy of the charged pion in the kaon rest
frame, and $W^2 = (q p_K)(q p_{\pm})/(m_{\pi^+}^2 m_K^2)$.  Factorizing the
IB differential width, one can write \cite{GINO}
\begin{eqnarray}
\Frac{\partial^2 \Gamma}{\partial T_c^* \partial W^2} \, & = & \, 
\Frac{\partial^2 \Gamma_{IB}}{\partial T_c^* \partial W^2} \, 
\left\{ \, 1 \, + \, 2 \, \Frac{m_{\pi^+}^2}{m_K} \, 
\Re \left( \Frac{E_{DE}}{e A} \right) \, W^2 \, \right. \no \\
& & \qquad \qquad \left. + \, \Frac{m_{\pi^+}^4}{m_K^2} \left( \left|
 \Frac{E_{DE}}{e A}
\right|^2 \, + \, \left| \Frac{M}{eA} \right|^2 \right) \, W^4 
\, \right\}~, \label{eq:ddtw2}
\end{eqnarray}
where $A \equiv A(K^{\pm} \rightarrow \pi^{\pm} \pi^0)$.  Since the linear
term in $W^2$ is sensitive to the interference between the IB amplitude and
the first electric dipole term $E_1$, it is convenient to introduce a
direct--$CP$--violating observable $\Omega$, defined as follows
\begin{equation}
\Frac{
{\partial^2 \Gamma^+}/{\partial T_c^* \partial W^2} 
\, - \, {\partial^2 \Gamma^- }/{\partial T_c^* \partial W^2} }{
{\partial^2 \Gamma^+}/{\partial T_c^* \partial W^2} 
\, + \, {\partial^2 \Gamma^- }/{\partial T_c^* \partial W^2} }
\; = \; 
\Omega \, W^2~\,+ \cO\left(\frac{m_\pi^4}{m_K^4}W^4\right) \, .
\label{eq:omw2}
\end{equation}

Setting $(p_1,p_2) \equiv (p_{\pm},p_0)$ and factorizing the strong phases
analogously to (\ref{eq:e1en}) we write \cite{DAI}, 
\beq 
E_1 (K^{\pm}) =
e^{i \delta_1}~E_c~, \qquad\qquad E_{IB}(K^{\pm}) = - e^{i \delta_2} ~
\Frac{3 e \Re(A_2)}{2 m_K z_{\pm} z_3} ~.
\label{eq:e1ib}
\eeq 
Assuming, as in the neutral channel, that the magnetic photon operator
gives the dominant SUSY contributions to the $CP$--violating phase of $E_c$,
we find \beq \Im (E_c)^\SUSY = \Im (E_n)^\SUSY ~, \eeq where $\Im
(E_n)^\SUSY$ is given in (\ref{eq:imen}).  Substituting this result in
(\ref{eq:ddtw2}) we finally obtain
\begin{eqnarray}
\Omega^\SUSY  & = &  \Frac{64}{99} \, \Frac{|\epsilon|}{\omega^2}
\Frac{m_s + m_d}{m_K} \, \sin(\delta_1 - \delta_2) \, \eta \,
\Frac{B_T}{B_G} \no \\
& & \, \times \, \left[ \Frac{F_0(x_{gq})}{G_0(x_{gq})}
\, + \, 8 ( 1 - \eta^{-1}) \, \right] \, \Re\left( \epop \right)^\SUSY_G~.
\label{eq:omiga} 
\end{eqnarray}
Since the dominant $CP$--conserving $K^\pm\to\pi^\pm\pi^0\gamma$ amplitude
is a $\Delta I = 3/2$ transition, $\Omega$ is enhanced by a factor
$\omega^{-2}$ with respect to $\epsp$.  This enhancement, however, is
partially compensated by the fact that the strong phase-difference
appearing in (\ref{eq:omiga}) is quite small $(\delta_1 - \delta_2)\simeq
10^\circ$ \cite{GASME}.\footnote{~While $\delta_2(m_K)\simeq - 7^{\circ}$
\cite{GASME}, in principle the $\delta_1$ phase shift should be input with
a dependence in the integration variables. This is however beyond the
accuracy required by the present analysis.}  Employing the same assumptions
adopted in Eq.~(\ref{epgbound}) and using $\sin(\delta_1 - \delta_2)\leq
0.2$ we find 
\beq 
|\Omega|^\SUSY~\leq~0.077~\Re\left( \epop \right)^\SUSY_G
\leq 2.3 \times 10^{-4}~.
\label{omigasusy}
\eeq 
Similarly to the case of $(\epsilon'_{+-\gamma}/\eps)^\SUSY$, also the
result in (\ref{omigasusy}) is substantially larger than what expected
within the Standard Model \cite{DAI}.\footnote{~An asymmetry at the level
of $10^{-4}$ between $K^+ \rightarrow \pi^+ \pi^0 \gamma$ and $K^-
\rightarrow \pi^- \pi^0 \gamma$ widths was claimed in \cite{DP} already
within the Standard Model. This result was however clearly overestimated as
discussed in \cite{DAI,RPS}.}

Since the kinetic variable $W^2$ can reach values of $\cO(1)$ \cite{DMS},
the result (\ref{omigasusy}) implies that in a specific region of the
Dalitz plot, the asymmetry between $K^+ \rightarrow \pi^+ \pi^0 \gamma$ and
$K^- \rightarrow \pi^- \pi^0 \gamma$ distributions can be of
$\cO(10^{-4})$.  A much smaller value is obtained performing a wide
integration over the phase space.  For instance integrating over $W$ and
$T_c^*$ in the interval $55~\mbox{MeV} \leq T_c^* \leq 90 ~\mbox{MeV}$
\cite{PDG}, leads to
\begin{equation}
\delta \Gamma = \Frac{\Gamma(K^+ \rightarrow \pi^+ \pi^0 \gamma) \, - \,
\Gamma(K^- \rightarrow \pi^- \pi^0 \gamma)}{\Gamma(K^+ \rightarrow \pi^+
\pi^0 \gamma) \, + \, \Gamma(K^- \rightarrow \pi^- \pi^0 \gamma)} \leq
3\times 10^{-3}~ \Re\left( \epop \right)^\SUSY_G~.
\label{eq:dgove}
\end{equation}

As pointed out in \cite{DP}, we finally note that $CPT$ invariance allows us
to connect, at the first order in $\alpha_{em}$, the charge asymmetry of
the total widths in $K^\pm \rightarrow \pi^\pm \pi^0 \gamma$ to the one in
$K^\pm \rightarrow \pi^\pm \pi^0$. The relation is given by
\begin{eqnarray}
\Frac{\Gamma(K^+ \rightarrow \pi^+ \pi^0) \, - \, 
\Gamma(K^- \rightarrow \pi^- \pi^0)}{\Gamma(K^+ \rightarrow \pi^+ \pi^0)
\, + \, \Gamma(K^- \rightarrow \pi^- \pi^0)} \, &  = & \,
- \, \Frac{B(K^+ \rightarrow \pi^+ \pi^0 \gamma)}{B(K^+ \rightarrow 
\pi^+ \pi^0)} \, \delta \Gamma \no \\
& \simeq & \, - 1.3 \times 10^{-3} \, \delta \Gamma~.
\label{eq:kppasy}
\end{eqnarray}
that, through (\ref{eq:dgove}), leads to an asymmetry of $\cO(10^{-8})$ for
the non--radiative process.

\section{Conclusions}
The unexpectedly large values of $\Re(\epsilon'/\epsilon)$ recently put
forward by the KTeV and the NA48 collaborations need a better theoretical
understanding. The difference from most SM estimates could be explained
either with unknown (but standard) non--perturbative effects or with New
Physics. Since the theoretical improvements in the calculation of the
non--perturbative effects may require a long time, it is worth looking for
other observables that could confirm or exclude the New--Physics origin of
the observed direct $CP$ violation.
\par
In this letter we have pointed out a strict correlation between the SUSY
contributions to the chromomagnetic operator, affecting
$\epsilon'/\epsilon$, and the magnetic $s d \gamma$
operator contributing to $K\to\pi\pi\gamma$ amplitudes.
We have searched for direct--$CP$--violating observables in the latter
processes which may get enhanced by a large coefficient in front of the
magnetic--penguin operator.
\par
First we have considered $K_{L,S} \rightarrow \pi^+ \pi^- \gamma$ decays
and concluded that the ratio $\epsilon'_{+-\gamma}/\epsilon$ is presumably
enhanced over its SM value in the scenario where $\epsilon'/\epsilon$ is
dominated by gluino--mediated supersymmetric amplitudes.  In particular for
large photon energies $|\epsilon'_{+-\gamma}/\epsilon|$ could reach values
of $\cO(0.5\%)$.  In the $K^{\pm} \rightarrow \pi^{\pm} \pi^0
\gamma$ modes we have studied the charge asymmetry of the decay
distributions.  We have found that also this clean direct--$CP$--violating
observable could be enhanced by supersymmetric effects, reaching values of
$\cO(10^{-4})$ in specific phase--space regions.
\par
In both cases the results found imply that a more detailed experimental
investigation of $CP$ violation in $K\to \pi\pi\gamma$ decays is well worth
the effort. Interestingly, this investigation could already be started with
existing experimental facilities like KTeV, NA48 and KLOE.  Finally, we
stress that the major theoretical uncertainty in the present analysis comes 
from the ratio of hadronic matrix elements $B_T/B_G$. We hope that this
quantity could be pinned down more precisely in the future with lattice--QCD
calculations.

\section*{Acknowledgements}
We thank F.J. Botella and  A.J. Buras for interesting discussions.
J.P. is partially supported by Grants AEN--96/1718 of CICYT (Spain) and
PB97--1261 of DGESIC (Spain). 
G.C. is partially supported by Schweizerische Nationalfonds.

\end{document}